\documentclass[
twocolumn,
nofootinbib,
amsmath,amssymb,
aps,
prd,
]{revtex4-1}

\pdfoutput=1

\usepackage{graphicx}
\usepackage{dcolumn}
\usepackage{bm}
\usepackage{natbib}
\usepackage{tikz}

\usepackage{hyperref}

\newcommand{\der}{\mathrm{d}}
\newcommand{\mpl}{M_{\mathrm{Pl}}}
\def\checkmark{\tikz\fill[scale=0.4](0,.35) -- (.25,0) -- (1,.7) -- (.25,.15) -- cycle;}

\begin{document}


\title{Large-$\eta$ Constant-Roll Inflation Is Never An Attractor}

\author{Michael J. P. Morse}
 \email{mjmorse3@buffalo.edu}
\author{William H. Kinney}
 \email{whkinney@buffalo.edu}
\affiliation{
Dept. of Physics, University at Buffalo
}

\date{April 5, 2018}

\begin{abstract}
Slow roll solutions to inflationary potentials have been widely believed to be the only universal attractor. Over the last few years there has been growing interest in a new class of inflationary models known as {\it Constant-Roll Inflation}. Constant roll solutions are a generalization of ``ultra-slow roll'' dynamics, where the first slow roll parameter is small, but the second slow roll parameter $\eta$ is larger than unity. In Ultra-slow Roll Inflation, the large-$\eta$ solution is a dynamical transient, relaxing exponentially to the attractor de Sitter solution. In the constant roll generalization, recent papers have concluded that Constant-Roll Inflation represents a new class of non-slow roll attractor solutions. In this paper we show that these attractor solutions are actually the usual slow roll attractor, disguised by a parameter duality, and that the large-$\eta$ solutions, as in the case of ultra-slow roll, represent a dynamical transient.

\end{abstract}

\maketitle

\section{Introduction}
On of the most promising theories of the early universe is that of inflation \cite{Starobinsky:1980te,Sato:1981ds,Sato:1980yn,Kazanas:1980tx,Guth:1980zm,Linde:1981mu,Albrecht:1982wi}. Inflation predicts a period of accelerated expansion in the early universe, which provides simultaneous solutions for the flatness and homogeneity of the current universe. Furthermore, inflation predicts that scalar curvature perturbations from quantum fluctuations of the scalar field generating inflation result in areas of over- and under-density in the universe, leading to structure formation \cite{Starobinsky:1979ty,Mukhanov:1981xt,Mukhanov:2003xw,Linde:1983gd,Hawking:1982cz,Hawking:1982my,Starobinsky:1982ee,Guth:1982ec,Bardeen:1983qw}. One family of inflationary models is that of single scalar fields. In single-field inflation, the field responsible for inflation, or {\it inflaton}, is taken to be displaced from a potential minimum. The field evolution is responsible for the accelerated expansion in the early universe which ends when the field value gets close to the minimum. 

In general the equation of motion of the inflaton in a given potential is not analytically solvable, and approximations are necessary. One widely used approximation scheme is that of slow-roll inflation \cite{Albrecht:1982wi,Steinhardt:1984jj,Linde:1981mu}. The slow roll approximation allows for analytic solutions for a wide range of potentials. More recently, ultra-slow roll and the more general case of constant roll solutions have been studied \cite{Kinney:1997ne,Martin:2012pe,Inoue:2001zt,Kinney:2005vj,Namjoo:2012aa,Huang:2013lda,Mooij:2015yka,Cicciarella:2017nls,Akhshik:2015nfa,Scacco:2015spa,Barenboim:2016mmw,Cai:2016ngx,Odintsov:2017yud,Grain:2017dqa,Odintsov:2017qpp,Bravo:2017wyw,Bravo:2017gct,Dimopoulos:2017ged,Nojiri:2017qvx,Motohashi:2017vdc,Odintsov:2017hbk,Oikonomou:2017xik,Cicciarella:2017nls,Awad:2017ign,Anguelova:2017djf,Salvio:2017oyf,Yi:2017mxs,Cai:2017bxr,Mohammadi:2018oku,Gao:2018tdb,Gao:2018cpp,Anguelova:2018ntr,Mohammadi:2018wfk,Karam:2017rpw}.

In this paper we show that when the constant-roll (large $\eta$) solution is an attractor, it is due to a duality which maps it uniquely onto a slow-roll (small $\eta$) solution. This result can be attributed to a transformation that maps one branch solution and potential onto the other. This invariant transformation was discovered in Ref. \cite{Tzirakis:2007bf} in the study of ultra-slow roll solutions. We will extend this to the inflationary model studied in Ref. \cite{Motohashi:2014ppa} and show that the constant-roll ``attractor'' solution is in fact the usual slow roll attractor disguised by a parameter duality, and that constant roll dynamics is a dynamical transient. This phenomenon is hidden when the Hamiltonian-Jacobi formulation \cite{Muslimov:1990be,Salopek:1990jq,Lidsey:1995np} is used to reduce the second order field equation into a system of first order differential equations. This method picks one one solution branch while leaving the other unrealized. When the second dynamical solution is obtained, the transformation becomes apparent.

This paper is structured as follows: In Section \ref{sec:singleFieldInflation} we give a brief review of single-field inflation and the slow-roll approximations. In Section \ref{sec:CurvaturePerturbations} we discuss curvature perturbations and develop the mode equation which will be used in later sections. Section \ref{sec:USRReview} reviews the work done in Ref. \cite{Tzirakis:2007bf} and show the duality of the inflation branch solutions in a simple model. The main results of this paper are developed in Section \ref{sec:ConstantRoll} and Section \ref{sec:ParameterSpace} where it is shown that the constant roll solutions have the same duality discussed in section \ref{sec:USRReview}, and the constant roll parameter space is compared to Planck constraints \cite{Planck:2013jfk}. Section \ref{sec:conclusion} presents conclusions.
\section{Single-Field Inflation}
\label{sec:singleFieldInflation}
In this section we review the formulation of single-field inflation and the slow-roll approximation.

For a minimally coupled scalar field $\phi$, the action can be written as
\begin{align}
S = \int \der^4 x \sqrt{-g}\left\lbrace \mathcal{L}_{EH} - \frac{1}{2}g^{\mu \nu} \partial_\mu \phi \partial_\nu \phi - V(\phi) \right\rbrace,
\end{align}
where $V$ is the field potential and $\mathcal{L}_{EH}$ is the Einstein-Hilbert Lagrangian. In flat Friedmann-Lema\^{i}tre-Robertson-Walker (FLRW) space with metric
\begin{equation}
g_{\mu\nu} = dt^2 - a^2\left(t\right) d {\bf x}^2,
\end{equation}
the Friedmann equation and equation of motion for a homogeneous field mode take the form
\begin{align}
\label{eqn:FriedmannI}
3 \mpl^2 H^2 &= \frac{\dot{\phi}^2}{2} + V, \\
\label{eqn:FieldEOM}
\ddot{\phi} + 3H \dot{\phi} + \frac{\der V}{\der \phi} &= 0.
\end{align}
The slow roll approximation is the limit that 
\begin{align}
\epsilon \equiv \frac{\dot{\phi}^2}{2\mpl^2 H^2} \ll 1,
\label{eqn:firstSRCondition}\\
\eta \equiv \frac{\ddot{\phi}}{H\dot{\phi}} \ll 1,
\label{eqn:secondSRCondition}
\end{align}
such that Eq. (\ref{eqn:FieldEOM}) and Eq. (\ref{eqn:FriedmannI}) take the forms
\begin{align}
3H \dot{\phi} + \frac{\der V}{\der \phi} &\approx 0, \\
\frac{V(\phi)}{2\mpl^2} &\approx H ^2. 
\end{align}
The dimensionless parameters $\epsilon$ (\ref{eqn:firstSRCondition}) and $\eta$ (\ref{eqn:secondSRCondition}) are referred to as {\it slow-roll parameters}.

There has been a great deal of attention in the last few years to so-called {\it Constant-Roll Inflation}, where the second slow-roll condition is violated, and $\eta$ is taken to be a constant of order unity \cite{Kinney:1997ne,Martin:2012pe,Inoue:2001zt,Kinney:2005vj,Namjoo:2012aa,Cicciarella:2017nls,Akhshik:2015nfa,Scacco:2015spa,Barenboim:2016mmw,Cai:2016ngx,Odintsov:2017yud,Grain:2017dqa,Odintsov:2017qpp,Nojiri:2017qvx,Motohashi:2017vdc,Odintsov:2017hbk,Oikonomou:2017xik,Cicciarella:2017nls,Awad:2017ign,Anguelova:2017djf,Salvio:2017oyf,Yi:2017mxs,Cai:2017bxr,Mohammadi:2018oku,Gao:2018tdb,Gao:2018cpp,Anguelova:2018ntr,Mohammadi:2018wfk,Karam:2017rpw}. Constant roll dynamics is a generalization of the ``ultra-slow roll'' case, where $\eta = 3$, and the dynamics is a dynamical transient asymptotically approaching de Sitter evolution. \cite{Kinney:2005vj}.

In order to solve the field equations it is useful to assume that the field is monotonic and can be used as an effective clock. This is a reasonable assumption in the region where slow roll is valid and remains reasonable in constant roll as along as $\dot{\phi} \neq 0$ at any point. This assumption allows the use of the Hamiltonian-Jacobi formulation, where time derivatives can be expressed in terms of field derivatives. In this way time derivatives of the Hubble parameter are written as 
\begin{align}
\frac{\der H}{\der t} &= \frac{\der H}{\der \phi} \frac{\der \phi}{\der t}, 
\label{eqn:HJ}\\
\frac{\der^2 H}{\der t \der \phi} &= \frac{\der^2 H}{\der \phi^2} \frac{\der \phi}{\der t},
\label{eqn:D[HJ,phi]}
\end{align}
and
\begin{align}
-2 \mpl^2 H' &= \dot{\phi},
\label{eqn:FriedmannH'} \\
2[H'(\phi)]^2 - \frac{3}{\mpl^2}H^2(\phi) + \frac{1}{\mpl^4}V(\phi) &= 0.
\label{eqn:H'_H_Potential}
\end{align}
The two above first order coupled differential equations are equivalent to Eq. (\ref{eqn:FieldEOM}), the field equation of motion. Likewise, we can express the slow roll parameters as derivatives with respect to $\phi$,
\begin{eqnarray}
\epsilon &&= 2 \mpl^2 \left(\frac{H'\left(\phi\right)}{H\left(\phi\right)}\right)^2,\cr
\eta &&= 2 \mpl^2 \frac{H''\left(\phi\right)}{H\left(\phi\right)},\cr
\xi^2 &&= 4 \mpl^4 \frac{H'''\left(\phi\right) H'\left(\phi\right)}{H^2\left(\phi\right)}.
\end{eqnarray}

\section{Curvature Perturbations}
\label{sec:CurvaturePerturbations}
The evolution of metric curvature perturbations is governed by the Mukhanov-Sasaki equation \cite{Mukhanov:1985rz,Sasaki:1986hm}. On a comoving hypersurface ($\delta \phi = 0$), the mode equation can be written as
\begin{align}
v_k'' +\left(k^2 - \frac{z''}{z} \right)v_k =0,
\end{align}
where 
\begin{align}
z = a \frac{\dot{\phi}}{H},
\end{align}
and the gauge invariant mode is given by
\begin{align}
v_k = z \zeta_k,
\end{align}
where $\zeta_k$ is the metric curvature perturbation. A prime denotes a derivative with respect to conformal time. 

The $z''/z$ term in the mode equation can be exactly written in terms of the first three slow roll parameters as
\begin{align}
\frac{z''}{z} &= 2 a^2 H^2(1 + \epsilon + \epsilon^2 - \frac{3}{2}\eta + \frac{1}{2}\eta^2 - 2 \epsilon\eta + \frac{1}{2}\xi^2)  \\ \nonumber &=a^2H^2F(\epsilon,\eta,\xi).
\label{eqn:z''/z}
\end{align}
Defining a new variable 
\begin{align}
y = \frac{k}{aH},
\end{align}
the Mukhanov-Sasaki equation can be re-written in the form
\begin{equation}
y^2(1-\epsilon^2)\frac{\der^2 u_k}{\der y^2} + 2 y \epsilon(\epsilon - \eta)\frac{\der u_k}{\der y} + [y^2 - F(\epsilon,\eta,\xi)] u_k =0.
\label{eqn:Mukhanov-Sasaki}
\end{equation}
When the first slow roll condition ($\epsilon \ll 1$) holds, but $\eta$ is not small this equation simplifies to
\begin{align}
y^2\frac{\der^2 u_k}{\der y^2} + [y^2 - (2 - 3 \eta + \eta^2)] u_k =0.
\label{eqn:modeEquationSlowRoll}
\end{align}

\section{Review of Duality in Ultra-Slow Roll Inflation}
\label{sec:USRReview}
In this section we review and show the solution duality in the case of ultra-slow roll, where $\eta = 3$ \cite{Kinney:1997ne,Tsamis:2003px,Kinney:2005vj}, first noted in the case of ``over the hill'' transient solutions for inverted quadratic potentials \cite{Tzirakis:2007bf}, where it was realized that this duality exists not only in the classical field equation, but carries over to the perturbation mode equation as well. Here we consider for simplicity the duality in the case of ultra-slow roll, where $V(\phi) = V_0 = \mathrm{const.}$ 

Take a potential of the form
\begin{align}
V = V_0,
\end{align}
where $V_0$ is a constant. The inflaton's equation of motion can now be written as
\begin{align}
\ddot{\phi} &= -3 H \dot{\phi}, \\
\eta &= 3.
\label{eqn:EOM_USR}
\end{align} 
If we take the late-time de Sitter evolution the scale factor is approximately constant
\begin{align}
H \approx \sqrt{\frac{V_0}{3 \mpl^2}}.
\end{align}
Under these approximations the field equation can be solved for $\phi(t)$
\begin{align}
\phi(t)  = A e^{-3Ht} + Be^{(0) t},
\end{align}
where we have expressed the exact de Sitter Solution ($\dot{\phi} = 0$) in the form $ \phi\left(t\right) = Be^{(0)Ht}$. With the knowledge of what we are looking for we can write the solution in an even more revealing form, remembering that, $\eta = 3$: 
\begin{align}
\phi(t)  = A e^{-\eta Ht} + Be^{(3-\eta)Ht}.
\label{eqn:USRSOLUTION}
\end{align}
Where we have done nothing but rewrite the constants in terms of $\eta = 3$ in our solution. In this form it is easy to see that the solution is invariant under the transformation
\begin{align}
\tilde{\eta} = 3 - \eta.  
\label{eqn:etaTransform}
\end{align} 
The de Sitter attractor branch $\phi = \mathrm{const.}$, corresponding to the slow roll solution for a constant potential, and the ultra slow roll transient branch $\phi \propto e^{-3 H t}$ simply swap positions under this transformation, leaving the solution invariant.

The in the limit of small $\epsilon$, but not small $\eta$, scalar perturbations are governed by Eq. (\ref{eqn:modeEquationSlowRoll}), which is also invariant under the duality transformation (\ref{eqn:etaTransform}). Equation (\ref{eqn:modeEquationSlowRoll}) has Hankel function solutions
\begin{align}
u_k &\propto \sqrt{y} \left[ a_k H_\nu (y) + b_k H^*_{\nu}(y) \right], \\
\nu &= \eta - \frac{3}{2}.
\end{align}
To satisfy the Bunch-Davies boundary condition,
\begin{align}
u_k \propto e^{iy},
\end{align}
we set $b_k=0$ and $a_k = 1$. However, under the duality in $\eta$, $\nu \rightarrow -\nu$ and the mode solution becomes
\begin{align}
u_k &\propto \sqrt{y} \left[ \tilde{a}_k H_{-\nu} (y) + \tilde{b}_k H^*_{-\nu}(y) \right].
\end{align}
If we use the relation
\begin{align}
H_{-\nu} = e^{i\nu \pi}H_{\nu},
\end{align}
the mode solution is rewritten in terms of order $\nu$ Hankel functions as,
\begin{align}
u_k &\propto \sqrt{y} \left[ \tilde{a}_k  e^{i\nu \pi} H_{\nu} (y) + \tilde{b}_k e^{-i\nu \pi} H^*_{\nu}(y) \right]. \\
\end{align}
Now to satisfy the Bunch-Davies boundary condition $\tilde{b}_k = 0, \tilde{a}_k = e^{-i\nu \pi}$. i.e the transformation has introduced an irrelevant overall phase factor to the mode solution.  This result should come as no surprise since the mode equation and the classical solution both exhibit the same invariance \cite{Tzirakis:2007bf}.

In the next section we will show that the constant roll solutions derived in Ref. \cite{Motohashi:2014ppa} exhibit the same invariance under the duality transformation (\ref{eqn:etaTransform}).  

\section{Constant-Roll Solutions}
\label{sec:ConstantRoll}
Constant-Roll Inflation is a generalization of ultra-slow roll introduced in Ref. \cite{Martin:2012pe}, where the second slow-roll parameter is expressed as 
\begin{align}
\eta = - \frac{\ddot{\phi}}{H \dot{\phi}} = (3 + \alpha).
\label{eqn:alphaReproduce}
\end{align}  
It can be seen that the limit $\alpha \rightarrow -3$ is the usual slow roll limit and $\alpha \rightarrow 0$ is the ultra-slow roll limit. The goal of this section is to show that there exist two branches of solution symmetric under a dual transformation 
\begin{align}
\tilde{\alpha} = -(3 + \alpha),
\label{eqn:AlphaDualRelation}
\end{align} 
arising from the fact that we could have written Eq. (\ref{eqn:alphaReproduce}) in another way, as
\begin{equation}
\eta = - \frac{\ddot{\phi}}{H \dot{\phi}} = -\tilde{\alpha}
\label{eqn:alphaDualRepresentation}
\end{equation}
to obtain the second branch solution to the second order field equation (\ref{eqn:FieldEOM}), where $\tilde{\alpha}=0$ is the slow roll solution and $\tilde{\alpha}=-3$ is the ultra-slow roll solution. We then show that this symmetry in the classical equations of motion is carried over into the perturbation equations, and is in fact the same symmetry derived in Ref. \cite{Tzirakis:2007bf}. To derive the general constant-roll solution, we use the Hamilton-Jacobi formalism, following Section II of Ref. \cite{Motohashi:2014ppa}.

Taking the time derivative of Eq. (\ref{eqn:FriedmannH'}), we get an equation that relates $\ddot{\phi}$ and $H''$ which will be used in both of the following cases to construct the two branch solutions of $\phi(t)$
\begin{align}
-2\mpl^2 H''  = \frac{\ddot{\phi}}{\dot{\phi}} \; \;.
\label{H''ddotPhi}
\end{align}
\subsection{$\eta = (3+\alpha)$}
With the help of Eq. (\ref{eqn:alphaReproduce}) and Eq. (\ref{H''ddotPhi}), we are able to construct a differential equation in terms of the Hubble parameter,
\begin{align}
-H \eta &= -2\mpl^2 H'', \\
H'' &= \frac{3 + \alpha}{2 \mpl^2}H,
\label{eqn:Hdiff}
\end{align}
which has the general solution
\begin{align}
H(\phi) = &A \exp\left( \sqrt{\frac{\alpha + 3}{2}} \frac{\phi}{\mpl}\right) \nonumber \\ 
+ &B \exp\left( -\sqrt{\frac{\alpha+3}{2}} \frac{\phi}{\mpl}\right),
\label{eqn:H_General_3+a}
\end{align}
corresponding to a potential given by
\begin{eqnarray}
V(\phi) = &&(-\alpha) A^2 \exp\left( 2\sqrt{ \frac{\alpha + 3}{2}} \frac{\phi}{\mpl}\right) \nonumber \\
+ &&(-\alpha) B^2 \exp\left( -2 \sqrt{\frac{\alpha+3}{2}} \frac{\phi}{\mpl}\right)\cr \nonumber \\
+ &&2(3+\alpha)AB + 6AB. 
\end{eqnarray}
We are, however, most interested in a special form of the solution
\begin{align}
H(\phi) &= H_0 \cosh\left( \sqrt{\frac{3 +\alpha }{2}} \frac{\phi}{\mpl}\right),
\label{eqn:HReproduce} \\
H(\phi)' &= \frac{H_0}{\mpl} \sqrt{\frac{3 +\alpha }{2}} \sinh\left( \sqrt{\frac{3 +\alpha }{2}} \frac{\phi}{\mpl}\right),
\label{eqn:HprimeReproduce}
\end{align}
corresponding to a generalization of the ultra-slow roll solution in Ref. \cite{Kinney:2005vj}. Under this definition of the Hubble parameter, the potential becomes 
\begin{align}
V(\phi) = \mpl^2 H_0^2 \times \left[3 \cosh\left( \sqrt{\frac{3 +\alpha }{2}} \frac{\phi}{\mpl}\right)^2 \right. \nonumber \\ -\left. (3 + \alpha)\sinh\left( \sqrt{\frac{3 +\alpha }{2}} \frac{\phi}{\mpl}\right)^2 \right].
\end{align}
Equation (\ref{eqn:FriedmannH'}) can now be used to find the field solution
\begin{align}
\frac{\phi}{\mpl} = \sqrt{\frac{2}{3+\alpha}} \mathrm{arctanh} \left[ \exp\left\lbrace -(3+\alpha) H_0 t\right\rbrace \right].
\label{eqn:phi(t)Reproduce}
\end{align}
Equation (\ref{eqn:phi(t)Reproduce}) is equivalent to Eq. (20) in Ref. \cite{Motohashi:2014ppa} under the definition of arctanh($2x$). For $\alpha < -3$ and $\alpha>0$, the hyperbolic trig functions switch into trig functions, so the solutions become
\begin{align}
H(\phi) &= H_0 \cos\left( \sqrt{\frac{|3 +\alpha|}{2}} \frac{\phi}{\mpl}\right),\\
\frac{\phi}{\mpl} &= \sqrt{\frac{2}{3+\alpha}} \mathrm{arctan} \left[ \exp\left\lbrace -(3+\alpha) H_0 t\right\rbrace \right], \\
V(\phi) &= \mpl^2 H_0^2 \left[3 \cos\left( \sqrt{\frac{|3 +\alpha|}{2}} \frac{\phi}{\mpl}\right)^2 \right. \nonumber \\ &\left. - (3 + \alpha)\sin\left( \sqrt{\frac{|3 +\alpha|}{2}} \frac{\phi}{\mpl}\right)^2 \right].
\end{align}
The solutions represent a branch of the $\phi(t)$ solution evolving on the given potential in their relevant $\alpha$ parameter regions.

\subsection{$\eta = -\alpha$}
We can construct the other $\phi(t)$ solution and the corresponding potential by using Eq. (\ref{eqn:alphaDualRepresentation}) as our definition of $\eta$.  The existence of this branch is recognized in Ref. \cite{Motohashi:2014ppa}: however, they do not derive it. As before, we use Eq. (\ref{eqn:alphaReproduce}) and Eq. (\ref{H''ddotPhi}) to construct a differential equation for the Hubble parameter.
\begin{align}
-H \eta &= -2\mpl^2 H'', \\
H'' &= -\frac{ \alpha}{2 \mpl}H.
\end{align}
This has general solution 
\begin{align}
H(\phi) &= A \exp\left( \sqrt{\frac{-\alpha}{2}} \frac{\phi}{\mpl}\right) \nonumber \\ &+ B \exp\left( -\sqrt{\frac{-\alpha}{2}} \frac{\phi}{\mpl}\right), 
\label{eqn:H_General_-a}
\end{align}
where to match with Eq. (\ref{eqn:phi(t)Reproduce}) with $(3+ \alpha >0)$ and $\alpha<0$, we write $\sqrt{\alpha} = i \sqrt{-\alpha}$, and write the solution as real exponentials instead of imaginary in order to match up with the branch solution (\ref{eqn:H_General_3+a}). The corresponding potential has the form
\begin{eqnarray}
V(\phi) = &&(3 + \alpha) A^2 \exp\left(2 \sqrt{ \frac{-\alpha}{2}} \frac{\phi}{\mpl}\right)  \nonumber \\ &&  + (3+\alpha) B^2 \exp\left( -2 \sqrt{\frac{-\alpha}{2}} \frac{\phi}{\mpl}\right)\cr \nonumber \\
&&+ 2(-\alpha)AB + 6AB. 
\end{eqnarray}
We are again interested in the particular case where 
\begin{align}
H(\phi) &= H_0 \cosh\left( \sqrt{\frac{-\alpha}{2}} \frac{\phi}{\mpl}\right), \\
H(\phi)' &=  \frac{H_0}{\mpl}\sqrt{\frac{-\alpha}{2}} \sinh\left( \sqrt{\frac{-\alpha}{2}} \frac{\phi}{\mpl}\right),
\end{align}
with a background potential of the form
\begin{align}
V(\phi) = \mpl^2 H_0^2 &\left[3 \cosh\left( \sqrt{\frac{-\alpha }{2}} \frac{\phi}{\mpl}\right)^2  \right. \nonumber \\ & \left.- (-\alpha)\sinh\left( \sqrt{\frac{-\alpha }{2}} \frac{\phi}{\mpl}\right)^2 \right].
\end{align}
Equation (\ref{eqn:FriedmannH'}) can again be used to find $\phi(t)$.
\begin{align}
\frac{\phi}{\mpl} = \sqrt{\frac{2}{-\alpha}} \mathrm{arctanh} \left[ \exp\left\lbrace \alpha H_0 t\right\rbrace \right].
\label{eqn:phi(t)Dual}
\end{align}

\subsection{Duality in constant roll solutions}
\label{sec:DualityConstantRoll}
The solution to the original second order field equation (\ref{eqn:FieldEOM}) has two branch solutions. In the region $-3<\alpha<0$, the solutions are Eq. (\ref{eqn:phi(t)Reproduce}) and Eq. (\ref{eqn:phi(t)Dual}):
\begin{align}
\frac{\phi_1(t)}{\mpl} &=  \sqrt{\frac{2}{3+\alpha}} \mathrm{arctanh} \left[ \exp\left\lbrace -(3+\alpha) H_0 t\right\rbrace \right], \\
\frac{\phi_2(t)}{\mpl} &=  \sqrt{\frac{2}{-\alpha}} \mathrm{arctanh} \left[ \exp\left\lbrace \alpha H_0 t\right\rbrace \right],
\label{eqn:phi(t)FullSolution}
\end{align}
evolving on potentials
\begin{align}
V_1(\phi)  &= \mpl^2 H_0^2 \left[3 \cosh\left( \sqrt{\frac{3 +\alpha }{2}} \frac{\phi}{\mpl}\right)^2 \right. \nonumber \\ &\left. -(3 + \alpha)\sinh\left( \sqrt{\frac{3 +\alpha }{2}} \frac{\phi}{\mpl}\right)^2 \right], \label{eq:V1a} \\
V_2(\phi) &= \mpl^2 H_0^2 \left[3 \cosh\left( \sqrt{\frac{-\alpha }{2}} \frac{\phi}{\mpl}\right)^2  \right. \nonumber \\ & \left.  +(\alpha)\sinh\left( \sqrt{\frac{-\alpha }{2}} \frac{\phi}{\mpl}\right)^2 \right].
\end{align}
For the regions $\alpha>0$ and $\alpha<-3$, the solution takes the form 
\begin{align}
\frac{\phi_1(t)}{\mpl} &=  \sqrt{\frac{2}{|3+\alpha|}} \mathrm{arctan} \left[ \exp\left\lbrace -(3+\alpha) H_0 t\right\rbrace \right], \\
\frac{\phi_2(t)}{\mpl} &=  \sqrt{\frac{2}{|\alpha|}} \mathrm{arctan} \left[ \exp\left\lbrace \alpha H_0 t\right\rbrace \right], 
\label{eqn:phi(t)FullSolutionNOTHYPERBOLIC} 
\end{align}
with potentials
\begin{align}
V_1(\phi)  &= \mpl^2 H_0^2 \left[3 \cos\left( \sqrt{\frac{|3 +\alpha|}{2}} \frac{\phi}{\mpl}\right)^2 \right. \nonumber \\ & \left. - (3 + \alpha)\sin\left( \sqrt{\frac{|3 +\alpha|}{2}} \frac{\phi}{\mpl}\right)^2 \right], \label{eq:V1b} \\
V_2(\phi) &= \mpl^2 H_0^2 \left[3 \cos\left( \sqrt{\frac{|\alpha| }{2}} \frac{\phi}{\mpl}\right)^2 \right. \nonumber \\ & \left.+ (\alpha)\sin\left( \sqrt{\frac{|\alpha| }{2}} \frac{\phi}{\mpl}\right)^2 \right].
\end{align}
With the potentials and field solutions written together it can be seen that the transformation $\alpha \rightarrow - \left(3 + \alpha\right)$ uniquely maps 
\begin{align}
\left\lbrace \phi_1, V_1 \right\rbrace \leftrightarrow \left\lbrace \phi_2, V_2 \right\rbrace, 
\end{align} 
such that $\alpha = -3$ maps to $\tilde\alpha = 0$, and $\alpha = 0$ maps to $\tilde\alpha = -3$. That is, the duality exchanges the constant-roll and slow-roll solutions.

This duality should actually come as no surprise, since the same duality was found in Ref. \cite{Tzirakis:2007bf} as we show in Sec. (\ref{sec:USRReview}). However, in Sec. (\ref{sec:USRReview}) the field equation is linearized by taking $H \approx$ Constant and $V = V_0 = \mathrm{Constant}$. This approximation results in a field equation that can be directly solved with a general solution of the form; $\phi = \phi_1 + \phi_2$. On this solution the duality is realized as a transformation $\phi_1 \rightleftharpoons \phi_2$. i.e mapping the branch solutions to each other while leaving the general solution for $\phi\left(t\right)$ invariant. In the more general case of constant roll, we have not taken either the Hubble parameter or the potential to be constant. Without these approximations the field equation is \textit{not} linear. As the equation is still second order it still has two branches, $\phi_1$ and $\phi_2$, each one resulting from a choice of how to represent $\eta$. In this case, the transformation manifests on the set of potential and solutions, mapping the branches onto each other. Therefore, the slow roll attractor solution $|\alpha| = |-\eta| <1$ has a second representation under the duality on the range $-4 < \alpha < -3$. It can easily be shown that $\alpha = -\eta$ therefore Eq. (\ref{eqn:etaTransform}) and Eq. (\ref{eqn:AlphaDualRelation}) are two different representations of the same duality.

The automorphism can, however, be recovered in the small-field limit as follows: In the limit $\left(\phi / \mpl\right) \rightarrow 0$, Eqs. (\ref{eq:V1a}) and (\ref{eq:V1b}) both become
\begin{equation}
V_1\left(\phi\right) \rightarrow 3 \mpl^2 H_0^2 \left[1 - \frac{1}{2} \left(\frac{\alpha \left(3 + \alpha\right)}{3}\right) \left(\frac{\phi}{\mpl}\right)^2 + \cdots\right], \label{eq:smallfieldV}
\end{equation}
which is manifestly invariant under the duality $\alpha \rightarrow - \left(3 + \alpha\right)$. We can then calculate the {\it potential} slow-roll parameter $\eta_V$ as
\begin{equation}
\eta_V = \mpl^2 \left(\frac{V''\left(\phi\right)}{V\left(\phi\right)}\right) \rightarrow - \frac{\alpha \left(3 + \alpha\right)}{3},
\end{equation}
which is likewise invariant under $\alpha \rightarrow - \left(3 + \alpha\right)$, unlike the Hubble slow roll parameter (\ref{eqn:alphaReproduce}). The potential slow roll parameter and the Hubble slow roll parameter become equivalent in the slow-roll limit $\alpha \rightarrow -3$,
\begin{equation}
\eta_V \rightarrow 3 + \alpha = \eta,
\end{equation}
as expected. Note that the small-field limit is the observationally relevant region, since inflation requires $\epsilon < 1$, the Planck constraint on the scalar spectral index $n_S - 1 \simeq 0.96$ requires $\epsilon \ll 1$, and therefore $\left(\phi / \mpl\right) \ll 1$. We then see that the potential (\ref{eq:smallfieldV}) is just the case of Hilltop Inflation (for $\alpha > 0$ or $\alpha < -3$), or Hybrid Inflation (for $-3 < \alpha < 0$), studied in Refs. \cite{Kinney:1997ne,Kinney:2005vj,Tzirakis:2007bf}, where it was shown that the constant-roll solution represents a dynamical transient, and the attractor solution is slow roll. 

In the next section, we study perturbations in the limit $\epsilon \ll 1$, and show that the this duality extends to the scalar mode equations as well as the background solution \cite{Kinney:2005vj,Tzirakis:2007bf}.

\subsection{Scalar Perturbations}
Scalar perturbations on both branches are governed by Eq. (\ref{eqn:modeEquationSlowRoll}). The goal is now to express $z''/z$ in terms of $\alpha$ in order to explicitly see that the invariance under Eq. (\ref{eqn:AlphaDualRelation}) holds. Expressing the first three slow roll parameters in terms of $\alpha$ yields
\begin{align}
\epsilon &= (3 + \alpha) \left(\mathrm{tanh}\left[ \sqrt{\frac{(3 +\alpha)}{2}} \frac{\phi}{\mpl}\right]\right)^2,\\
\eta &= (3 + \alpha),\\
\xi^2 &= (3 + \alpha)^2 \left(\mathrm{tanh}\left[ \sqrt{\frac{(3 +\alpha)}{2}} \frac{\phi}{\mpl}\right]\right)^2 \nonumber \\&= (3 + \alpha) \epsilon,
\end{align}
and substituting these definitions into Eq. (\ref{eqn:z''/z}) we arrive at
\begin{align}
\frac{z''}{z} = 2 a^2 H^2 \left( \frac{1}{2}(1 + \alpha)(2+\alpha) - \frac{1}{2}( 7 + 3\alpha) \epsilon + \epsilon^2 \right).
\end{align}
We can now see that in the limit of $\epsilon \ll 1$, the term $z''/z$ is symmetric under the transformation Eq. (\ref{eqn:AlphaDualRelation}),  
\begin{align}
\frac{z''}{z} = 2 a^2 H^2 \left( \frac{1}{2}(1 + \alpha)(2+\alpha)\right).
\end{align}
In the $\epsilon \ll 1$ limit we write Eq. (\ref{eqn:Mukhanov-Sasaki}) as
\begin{align}
y^2\frac{\der^2 u_k}{\der y^2} + [y^2 -  \frac{1}{2}(1 + \alpha)(2+\alpha)] u_k =0,
\end{align} 
with Hankel function solutions of the form
\begin{equation}
u_k \propto \sqrt{y}\left[ a_k H_{\nu}(y)  + b_k H^*_{\nu}(y) \right],
\end{equation}
where
\begin{equation}
\nu = \left(\alpha +\frac{3}{2}\right).
\end{equation}
To satisfy the Bunch-Davies boundary condition, we set $b_k = 0,a_k = 1$, so these modes are of the form
\begin{align}
u_k = e^{iy}.
\end{align}
We can, however, now transform $\nu$ under the duality transformation. This will lead to a mode solution of the form
\begin{align}
u_k \propto \sqrt{y}\left[ {\tilde a}_k H_{-\nu}(y)  + {\tilde b}_k H^*_{-\nu}(y) \right].
\end{align}
However, as
\begin{align}
H_{-\nu} = e^{i\nu \pi}H_{\nu},
\end{align}
the solution can be written as
\begin{align}
u_k \propto \sqrt{y}\left[ {\tilde a}_k e^{i\nu \pi}H_{\nu}(y)  + {\tilde b}
_k e^{i\nu \pi}H^*_{\nu} \right].
\end{align}
where now the Bunch-Davies Boundary condition is satisfied by setting ${\tilde b}_k = 0$ and ${\tilde a}_k = e^{-i\nu \pi}$. i.e the transformation has simply introduced an irrelevant overall phase factor to the mode solution. The fact that the solution is invariant under the transformation should come as no surprise, since the mode equation is also invariant.

\section{Classifying parameter regions in Constant Roll}
\label{sec:ParameterSpace}
We have parameterized our constant roll solutions by $\alpha$, but have not yet given an interpretation to the different regions of $-\alpha = \eta$. In this section we will discuss the physical modes described by different parameter regions. We will find that some regions are characterized as slow-roll inflation, some regions are much more exotic inflationary potentials, and some regions do not describe inflation at all, but rather different types of non-inflationary expansion. Figure (\ref{table:models}) gives an overview of the different regions and their duals. We also classify the regions of parameter space which contain models consistent with the Planck constraint of $n_s = .9606 \pm 0.0073$ and $r<0.11$ \cite{Planck:2013jfk}. An important note here is that, while the regions are listed twice for completeness, a region and its dual are two representations of the same portion of the parameter space. Regions 1 and 7 have no small-$\eta$ limit, and we do not consider them further here. We discuss below the physical interpretation of Regions 2-6.
\begin{table}
\begin{tabular}{|c|c|c|c|c|}
\hline 
Region & $-\eta = \alpha$ & Dual Region & $-\tilde{\eta}=\tilde{\alpha}$  & Agreement with Planck \\ 
\hline
1 & $<-4$ & 7 & $>1$   & X\\
\hline 
2 &  (-4,-3) & 6  &  (0,1)  & \checkmark \\
\hline 
3 &  (-3,-2) & 5 &  (-1,0)   & X\\
\hline 
4 &  (-2,-1) & 4 &  (-2,-1)  & X \\
\hline 
5 &  (-1,0) & 3 &  (-3,-2)  & X \\
\hline  
6 &  (0,1) & 2  &  (-4,-3)  & \checkmark\\
\hline
7 &   $>1$ & 1  & $<-4$     & X\\
\hline
\end{tabular} 
\caption{$\alpha$ parameter region and model interpretation}
\label{table:models}
\end{table}

\subsection{Hilltop Inflation (Regions 2,6)}

The potential in Regions 2 and 6 can be written as
\begin{align}
&V\left(\phi\right) = 3 H_0^2 \mpl^2 \times \nonumber \\ &\left\lbrace1 - \frac{3 + \alpha}{6}\left[1 - \cos\left(2 \sqrt \frac{\left\vert 3 + \alpha\right\vert}{2} \frac{\phi}{\mpl}\right) \right]\right\rbrace.
\end{align}
In the small-field limit, this becomes
\begin{equation}
V\left(\phi\right) \rightarrow 3 \mpl^2 H_0^2 \left[ 1 - \frac{1}{2} \left\vert \frac{\alpha \left(3 + \alpha\right)}{3}\right\vert \left(\frac{\phi}{\mpl}\right)^2 + \cdots\right],
\end{equation}
corresponding to slow-roll Hilltop or Natural Inflation, with a slightly red-tilted scalar spectral index as predicted by Planck, $n_S - 1 = 2 \eta < 0$ \cite{Boubekeur:2005zm,Freese:1990rb,Motohashi:2014ppa}. Reference \citep{Motohashi:2014ppa} writes this potential in terms of $(3 + \alpha)$, and arrives at with region 2 as the only viable region. In fact these regions are equivalent under the $\alpha$ duality. In the limit that $\alpha \rightarrow -3$ we recover ultra-slow roll inflation with $V_0 = 3 M^2 \mpl^2$. This model was is discussed in Sec.(\ref{sec:USRReview}) where we see the general solution is of the form  
\begin{align}
\phi(t) &\approx \exp\left\lbrace -3 Mt\right\rbrace  + C 
\end{align}
where $C$ is a constant. The general case was considered in Ref. \cite{Tzirakis:2007bf}, where it was shown directly that the constant roll solution represents a dynamical transient, with slow-roll inflation as the unique attractor solution.

\subsection{Hybrid Inflation (Regions 3, 5)}

In Regions 3 and 5 from Table \ref{table:models}, the potential reduces to:
\begin{align}
&V\left(\phi\right) =\times \nonumber \\ & 3 H_0^2 \mpl^2 \left\lbrace1 - \frac{3 + \alpha}{6}\left[\cosh\left(2 \sqrt{\frac{3 + \alpha}{2}} \frac{\phi}{\mpl}\right) - 1 \right]\right\rbrace.
\end{align}
In the small-field limit, this becomes
\begin{equation}
V\left(\phi\right) \rightarrow 3 \mpl^2 H_0^2 \left[ 1 + \frac{1}{2} \left\vert \frac{\alpha \left(3 + \alpha\right)}{3}\right\vert \left(\frac{\phi}{\mpl}\right)^2 + \cdots\right],
\end{equation}
which we recognize as a slow-roll quadratic Hybrid inflation model \cite{Linde:1993cn}. As in the Hilltop case, we recover the Ultra-Slow Roll limit when $\alpha \rightarrow - 3$, and
\begin{equation}
V\left(\phi\right) \rightarrow 3 \mpl^2 H_0^2.
\end{equation}
This parameter region predicts a blue spectrum, $n_S - 1 = 2 \eta > 0$, and is inconsistent with current constraints from Planck. This case was considered in Ref. \cite{Kinney:2005vj}, where it was shown directly that the constant roll solution represents a dynamical transient. The attractor in this case is eternal inflation, with the field at the minimum of the potential and $\dot\phi = 0$, corresponding to the ``slow roll'' solution $\dot\phi = V'\left(\phi\right) / 3 H \rightarrow 0$.

\subsection{The Self-Dual Case (Region 4)}

The self dual Region 4 is not physically viable as an inflation model, since the scalar spectral index is not red tilted. However it is of interest to investigate, as it produces some unexpected models, including models where inflation does not occur at all. We are able to construct both radiation- and matter-dominated expansion by taking 
\begin{align}
H(\phi) = H_0 \exp\left\lbrace \sqrt{\frac{-\alpha}{2}} \frac{\phi}{\mpl}\right\rbrace
\end{align}
Where we have chosen to work with Eq. (\ref{eqn:H_General_-a}). This form of the Hubble parameter will lead to a potential of the form
\begin{align}
V(\phi) =  \mpl^2 ( 3 + \alpha) H_0^2   \exp\left\lbrace 2\sqrt{\frac{-\alpha}{2}} \frac{\phi}{\mpl}\right\rbrace
\end{align}
Which can be identified as the power-law potential \cite{Abbott:1984fp,Lucchin:1984yf}. where the scale factor grows as
\begin{align}
a(t) \propto t^{-\frac{1}{\alpha}}.
\end{align}
For the mid point solution, that is dual to its self, $\alpha = -\frac{3}{2}$ we see we get expansion of the form
\begin{align}
a(t) \propto t^{\frac{2}{3}}.
\end{align}  
which can be identified as expansion under matter domination.

Similarly, the end point of the region $\alpha = -2$ leads to expansion of governed by radiation domination
\begin{align}
a(t) \propto t^{\frac{1}{2}}.
\end{align} 
The other endpoint $\alpha = -1$ is the same solution under the duality $\alpha \rightarrow -(3 + \alpha)$.

\section{Conclusion}
\label{sec:conclusion}

We have considered a transformation present in constant-roll inflationary solutions which disguises slow-roll solutions. Since the equation of motion of a minimally coupled inflaton field is a second-order differential equation, it will have two linearly independent branch solutions. When both branches of the field solution and potentials are considered,
\begin{align}
\frac{\phi_1(t)}{\mpl} &=  \sqrt{\frac{2}{3+\alpha}} \mathrm{arctanh} \left[ \exp\left\lbrace -(3+\alpha) H_0t\right\rbrace \right],\\
V_1(\phi)  &= \mpl^2H_0^2 \left[3 \cosh\left( \sqrt{\frac{3 +\alpha }{2}} \frac{\phi}{\mpl}\right)^2 \right. \nonumber \\ & \left.- (3 + \alpha)\sinh\left( \sqrt{\frac{3 +\alpha }{2}} \frac{\phi}{\mpl}\right)^2 \right], \\
\frac{\phi_2(t)}{\mpl} &=  \sqrt{\frac{2}{-\alpha}} \mathrm{arctanh} \left[ \exp\left\lbrace \alpha H_0t\right\rbrace \right],\\
V_2(\phi) &= \mpl^2H_0^2 \left[3 \cosh\left( \sqrt{\frac{-\alpha }{2}} \frac{\phi}{\mpl}\right)^2 \right. \nonumber \\ & \left.+ (\alpha)\sinh\left( \sqrt{\frac{-\alpha }{2}} \frac{\phi}{\mpl}\right)^2 \right] 
\end{align} 
it can be seen that the solution is invariant under the transformation
\begin{align}
\alpha \rightarrow -(3+ \alpha),
\end{align}
which defines an isomorphism between the two branches. This transformation was discovered in Ref. \cite{Tzirakis:2007bf} for the case of ultra-slow roll inflation where the classical inflationary field equation of motion was solved without the Hamiltonian-Jacobi method, leading directly to the two branches. 

In previous literature on constant-roll solutions only one of the branches is constructed leading to the incorrect conclusion that the constant-roll branch solution is a new class of attractor solutions. However, when the duality is considered it is apparent that the attractor solution is still that of slow roll, just in disguise.

We show that in the limit $\epsilon \ll 1$ the duality symmetry is preserved in the curvature perturbation mode equation
\begin{align}
y^2\frac{\der^2 u_k}{\der y^2} + [y^2 - (2 + 3 \alpha + \alpha^2)] u_k =0,
\end{align}
and that in general the duality will result in at most an overall phase shift in the Hankel function solutions.
We confirm that this transformation preserves the solution given in Section III.A of Ref. \cite{Motohashi:2014ppa} without the need for a phase shift as for the Hankel function, $H_\nu$, where they define
\begin{align}
\nu = \left\vert \alpha + \frac{3}{2} \right\vert,
\end{align}
which is invariant under the transformation.

We conclude that slow roll remains the universal attractor solution, and that when the constant roll solutions appear to exhibit attractor behavior, they are actually just slow roll, disguised by the duality.

\section*{Acknowledgments}

This work is supported by the National Science Foundation under grants NSF-PHY-1417317 and NSF-PHY-1719690.

\bibliographystyle{apsrev4-1}
\bibliography{paper}

\end{document}